# Augmented Set-membership Affine Projection Algorithm and Its Performance Analysis

Xinnian Guo[1], Haiquan Zhao[2]*, Chen Wang[2], Xiaoqiang Long[2], Yalin Liu[2], Wenjing Luo[2]

Abstract: The augmented affine projection algorithm (AAPA) has considerably excellent performance for highly colored input signals. However, the direct matrix inversion operation leads to a high computational complexity, especially with high projection order. Inspired by the excellent characteristics of set-membership filtering (SMF), this paper proposes the augmented set-membership affine projection algorithm (ASM-APA), which not only has low computational complexity but also offers improved performance compared with AAPA. Then, the computational complexity and stability of ASM-APA are analyzed, and the condition for maintaining the stability of the algorithm **is** provided. Finally, in the computer simulation phase, the results of the simulation experiments demonstrated that ASM-APA has superior performance compared to AAPA.

## 1 Introduction

Adaptive filter has been extensively applied in active noise control (ANC) [1], system identification and echo cancellation [2-14]. The least mean square (LMS) algorithm and its normalized version (NLMS) are the most popular adaptive algorithms since their simple structure and easy implementation [15]. In several practical applications, the highly colored signals are not rare, such as the speech signal and wind speed signal. For such inputs, the classical LMS and NLMS algorithms converge slowly. Fortunately, the affine projection algorithm (APA) has gained a lot of attention as a good alternative algorithm that can greatly speed up the convergence of the algorithm [16]. However, the weight update of APA requires matrix inversion, which leads to its higher computational complexity. This drawback is more significant as the projection order increases. To overcome this drawback, the idea of set-membership (SM) filtering was introduced into the adaptive filtering framework, resulting in algorithms such as SM-NLMS and SM-APA [17–18], which both achieve better steady-state performance while reducing the complexity of their parent algorithms, i.e., NLMS and APA. Specifically, the SM filtering strategy updates the filter weights only when the magnitude of the estimation error is greater than a predetermined error bound, i.e., only the most informative data is eligible to drive filter updates. In fact, SM-APA is a generalization of the SM-NLMS algorithm [19]. In addition, unlike the APA, the SM-APA does not require the trade-off between the convergence rate and steady-state misalignment. Based on the energy conservation method, the authors in [20] analyzed the steady-state mean square error (MSE) of the SM-APA. In recent years, some variants of SM-APA have been widely investigated to solve problems encountered in practical applications, e.g., reducing complexity and identifying sparse systems [21,22].

However, it is worth mentioning that all the aforementioned SM-type algorithms are real-valued adaptive filtering algorithms for processing real signals. In fact, the complex signals are widely used in the fields of radar, sonar, power and communication [23]. For example, the frequency estimation in power systems [24], direction of arrival estimation [25], wind speed prediction [26], and so on. Among the classical complex-valued adaptive filtering algorithms, the complex-valued LMS (CLMS) algorithm is the most popular, which is a direct extension of the real-valued LMS algorithm in the complex domain with the same advantages. However, the CLMS algorithm can only achieve optimal estimation for circular inputs. The complex-valued signals can be categorized into second-order circular and non-circular signals based on the second-order statistics, which depends on the absence or presence of pseudo-covariance (The pseudo-covariance is defined as $\boldsymbol{P} = E\{\boldsymbol{x}\boldsymbol{x}^T\}$, where $\boldsymbol{x} \in \mathbb{C}^{L\times 1}$ is a complex-valued vector with length $L$). The pseudo-covariance of the non-circular signal is not zero, i.e., $\boldsymbol{P} \neq 0$ [27]. In order to better deal with the non-circular inputs, the CLMS algorithm based on the widely-linear (WL) model was proposed, which fully takes into account the second-order statistical information of the input signals and achieves better estimation performance than the CLMS algorithm under the non-circular inputs [18]. However, when the input signals are highly correlated, the convergence speed of the widely-linear complex-valued LMS (WL-CLMS) algorithm is severely degraded [27]. To this end, a WL model-based APA, called augmented APA (AAPA), was proposed to significantly improve the convergence speed of the WL-CLMS algorithm. In addition, its steady-state MSE was analyzed in [27]. Nevertheless, similar to real-valued APA, the AAPA suffers from high computational complexity.

---

[1] Suqian Key Laboratory of Visual Inspection and Intelligent Control, Suqian University, Suqian, 223800, China

[2] Key Laboratory of Magnetic Suspension Technology and Maglev Vehicle, Ministry of Education, School of Electrical Engineering, Southwest Jiaotong University, Chengdu 610031, China.
* Corresponding author: Haiquan Zhao (hqzhao_swjtu@126.com).
Xinnian Guo(xinnianguo@squ.edu.cn); Chen Wang (cc1005805@126.com); Xiaoqiang Long (xiaoqiangL@my.swjtu.edu.cn); Yalin Liu (lyl959946@163.com); Wenjing Luo (luowenjing0908@163.com).

Inspired by the merits of SM filtering, this paper proposes an augmented set-membership affine projection algorithm (ASM-APA), for which, to the best of our knowledge, no similar research has emerged. It is foreseeable that the proposed ASM-APA compare to AAPA can achieve significant performance improvement compared to AAPA with lower complexity, this will be confirmed in the computer simulation section. Then, we analyze the stability and computational complexity of ASM-APA and obtain the stability conditions. Numerous computer simulations, including system identification and SAEC, demonstrate the performance advantages of the proposed ASM-APA.

This paper is organized as follows: In Section 2, we briefly review SM filtering and AAPA. Section 3 presents the detailed derivation of the ASM-APA algorithm. The computational complexity and stability analysis of the F-algorithm is described in Section 4. Section 5 performs extensive computer simulations to demonstrate the superiority of the proposed F-algorithm. Finally, Section 6 summarizes the full text.

*Notation*: $E[\cdot]$ indicates expectation operator, $(\cdot)^T$ denotes transpose operator, $\mathrm{sgn}(\cdot)$ represents signum function. $\|\cdot\|$ is norm of vector or matrix, $\boldsymbol{I}$ stands for identity matrix with suitable dimension, $|\cdot|$ denotes the modulus of a complex number. respectively.

## 2. Review of the set-membership filtering and AAPA

### 2.1. Set-membership filtering

Define a set $\mathrm{S}$ containing all possible input-desired signal data pairs $\{\boldsymbol{x}, d\}$, and $\Phi$ (feasibility set) represents the set of filter weight vectors satisfying $\left|d - \boldsymbol{w}^T\boldsymbol{x}\right| \le \alpha$, i.e.,

$$\Phi = \bigcap_{(\boldsymbol{x},d)\in \mathrm{S}} \left\{\boldsymbol{w} \in \mathbb{C}^N ; \left|d - \boldsymbol{w}^T\boldsymbol{x}\right| \le \alpha\right\} \tag{1}$$

where $\alpha$ is a predefined error bound for constraining the hyperplane composed of filter weight vectors. At time instant *n*, the input-desired data pair $\{\boldsymbol{x}(n), d(n)\}$ is available and its corresponding constraint set $\mathcal{H}_n$ can be defined as

$$\mathcal{H}_n = \left\{ \boldsymbol{w} \in \mathbb{C}^N ; \left| d(n) - \boldsymbol{x}(n)^T \boldsymbol{w} \right| \le \alpha \right\} \tag{2}$$

The membership set is given by

$$\Psi_n = \bigcap_{i=1}^{n} \mathcal{H}_i \tag{3}$$

which implies the intersection of the constraint sets corresponding to the input-desired data pairs available from the initial time to the current time *n*. When *n* tends to infinity, the membership set $\Psi_n$ is equivalent to the feasible set $\Phi$, i.e., $\lim_{n\to\infty} \Psi_n \approx \Phi$ .

## 2.2 Augmented Affine Projection Algorithm

For the processing of second-order non-circular input signals, the widely-linear (WL) model can fully exploit their second-order statistical information. The output signal of the WL adaptive filter can be expressed as

$$y(n) = \boldsymbol{x}^T(n)\boldsymbol{h}(n) + \boldsymbol{x}^H(n)\boldsymbol{g}(n) \tag{4}$$

where $\boldsymbol{h}(n)$ and $\boldsymbol{g}(n)$ are the standard and conjugate filter weight vectors with length of *N*, respectively. $\mathbf{x}(n) = [x(n), x(n-1), ..., x(n-N+1)]^T$ is the input signal vector of length $N$ . In the context of system identification, the desired signal $d(n)$ of the WL filter is given by

$$d(n) = \boldsymbol{x}^T(n)\boldsymbol{h}_o + \boldsymbol{x}^H(n)\boldsymbol{g}_o + v(n) \tag{5}$$

where $\boldsymbol{h}_o$ and $\boldsymbol{g}_o$ are the standard and conjugate system parameter vectors, $v(n)$ is the background noise.

To improve the convergence performance of the augmented complex-valued LMS (ACLMS) algorithm under colored input signals, AAPA was proposed in [19]. By solving a constrained minimum perturbation problem using the Lagrange multiplier method and Cauchy-Riemann ( $\mathbb{CR}$ ) calculus, the weight update equation for AAPA is obtained as follows:

$$\boldsymbol{h}(n+1) = \boldsymbol{h}(n) + \mu \boldsymbol{X}^*(n)[\boldsymbol{X}^H(n)\boldsymbol{X}(n) + \boldsymbol{X}^T(n)\boldsymbol{X}^*(n) + \delta \boldsymbol{I}]^{-1} e(n) \tag{6}$$

$$\boldsymbol{g}(n+1) = \boldsymbol{g}(n) + \mu \boldsymbol{X}(n)[\boldsymbol{X}^H(n)\boldsymbol{X}(n) + \boldsymbol{X}^T(n)\boldsymbol{X}^*(n) + \delta \boldsymbol{I}]^{-1} e(n) \tag{7}$$

where $\boldsymbol{X}(n) = [\boldsymbol{x}(n), \boldsymbol{x}(n-1), \ldots, \boldsymbol{x}(n-P+1)]$ is the input signal matrix of size $N \times P$ , and $P$ is the projection order. $\mu$ is the step size, and $\delta$ is a small regularization parameter. The estimation error vector $e(n)$ is computed by

$$e(n) = \boldsymbol{d}(n) - \boldsymbol{X}^T(n)\boldsymbol{h}(n) - \boldsymbol{X}^H(n)\boldsymbol{g}(n) \tag{8}$$

where $\boldsymbol{d}(n) = [d(n), d(n-1), ..., d(n-P+1)]^T$ denotes the desired signal vector of length $P$ .

# 3. The proposed ASM-APA

In this section, a new low-complexity implementation of the augmented complex-valued adaptive filtering algorithm is proposed, called ASM-APA. Before deriving the ASM-APA algorithm in detail, some new definitions need to be provided. For the WL filter, the new feasibility set $\Phi$ and constraint set $\mathcal{H}_n$ are respectively redefined as

$$\Phi = \bigcap_{(\mathbf{x},d)\in S} \left\{ \boldsymbol{h}, \boldsymbol{g} \in \mathbb{C}^N ; \left| d - \boldsymbol{x}^T \boldsymbol{h} - \boldsymbol{x}^H \boldsymbol{g} \right|^2 \leq \alpha^2 \right\} \tag{9}$$

$$\mathcal{H}_n = \left\{ \boldsymbol{h}, \boldsymbol{g} \in \mathbb{C}^N ; \left| d(n) - \boldsymbol{x}^T(n)\boldsymbol{h}(n) - \boldsymbol{x}^H(n)\boldsymbol{g}(n) \right|^2 \leq \alpha^2 \right\} \tag{10}$$

The membership set $\Psi_n$ can be expressed as

$$\Psi_n = \bigcap_{i=1}^{n-P} \mathcal{H}_i \bigcap_{j=n-P+1}^{n} \mathcal{H}_j = \Psi_n^{n-P} \cap \Psi_n^P \tag{11}$$

where $\Psi_n^{n-P}$ is the intersection of the first $n-P$ constraint sets and $\Psi_n^P$ represents the intersection of the last $P$ constraint sets. In our proposed ASM-APA, our goal is to update $\boldsymbol{h}$ and $\boldsymbol{g}$ so that they lie in the set $\Psi_n^P$, with minimal weight perturbation, i.e., $\boldsymbol{h}_n, \boldsymbol{g}_n \in \Psi_n^P$.

Defined a complex-valued parameter space $S_{n-l+1}$ that contains all weight vectors $\boldsymbol{h}$, $\boldsymbol{g}$ that satisfy the following conditions, i.e.,

$$d(n-l+1) - \boldsymbol{x}^T(n-l+1)\boldsymbol{h} - \boldsymbol{x}^H(n-l+1)\boldsymbol{g} = c(n-l+1), l = 1,2,...,P \tag{12}$$

Note that if any selected $c(n-l+1)$ satisfies $|c(n-l+1)| \leq \alpha$, then $S_{n-l+1} \in \mathcal{H}_{n-l+1}$. In the light of the principle of minimum disturbance, the constraint minimization problem of two weight vectors is solved when $\boldsymbol{h}(n),\ \boldsymbol{g}(n) \notin \Psi_n^P$.

$$\begin{aligned} \min \quad & \|\boldsymbol{h}(n+1) - \boldsymbol{h}(n)\|^2 + \|\boldsymbol{g}(n+1) - \boldsymbol{g}(n)\|^2 \\ \text{s.t.} \quad & \boldsymbol{d}(n) - \boldsymbol{X}^T(n)\boldsymbol{h}(n+1) - \boldsymbol{X}^H(n)\boldsymbol{g}(n+1) = \boldsymbol{c}(n) \end{aligned} \tag{13}$$

where $\boldsymbol{c}(n) = [c_1(n), c_2(n), ..., c_P(n)]^T \in \mathbb{C}^P$ specifies the point in $\Psi_n^P$. Using the Lagrange multiplier method, the unconstrained optimization cost function is constructed as

$$\begin{aligned} J_{ASM-APA} = & \|\boldsymbol{h}(n+1) - \boldsymbol{h}(n)\|^2 + \|\boldsymbol{g}(n+1) - \boldsymbol{g}(n)\|^2 \\ & + \Re\left\{ [\boldsymbol{d}(n) - \boldsymbol{X}^T(n)\boldsymbol{h}(n+1) - \boldsymbol{X}^H(n)\boldsymbol{g}(n+1) - \boldsymbol{c}(n+1)]^H \cdot \boldsymbol{\beta}(n) \right\} \end{aligned} \tag{14}$$

where $\boldsymbol{\beta}(n)$ is a Lagrange multiplier vector.

Based on the $\mathbb{CR}$ calculus, taking the partial derivative of $J_{ASM-APA}$ with respect to $\boldsymbol{g}^*(n+1)$ yields $\dfrac{\partial J_{ASM-APA}(n)}{\boldsymbol{g}^*(n+1)} = \dfrac{1}{2}\left( \dfrac{\partial J_{ASM-APA}(n)}{\boldsymbol{g}_r(n+1)} + j\dfrac{\partial J_{ASM-APA}(n)}{\boldsymbol{g}_i(n+1)} \right) = \boldsymbol{g}(n+1) - \boldsymbol{g}(n) - \dfrac{1}{2}\boldsymbol{X}(n)\boldsymbol{\beta}(n)$, and make its derivative zero. The same reasoning leads to

$$\boldsymbol{h}(n+1) - \boldsymbol{h}(n) = \frac{1}{2}\boldsymbol{X}^*(n)\boldsymbol{\beta}(n) \tag{15}$$

$$\boldsymbol{g}(n+1) - \boldsymbol{g}(n) = \frac{1}{2}\boldsymbol{X}(n)\boldsymbol{\beta}(n) \tag{16}$$

Next, pre-multiplying both sides of (15) and (16) by $\boldsymbol{X}^T(n)$ and $\boldsymbol{X}^H(n)$ respectively yields

$$\boldsymbol{X}^T(n)\boldsymbol{h}(n+1)-\boldsymbol{X}^T(n)\boldsymbol{h}(n)=\frac{1}{2}\boldsymbol{X}^T(n)\boldsymbol{X}^*(n)\boldsymbol{\beta}(n) \tag{17}$$

$$\boldsymbol{X}^H(n)\boldsymbol{g}(n+1)-\boldsymbol{X}^H(n)\boldsymbol{g}(n)=\frac{1}{2}\boldsymbol{X}^H(n)\boldsymbol{X}(n)\boldsymbol{\beta}(n) \tag{18}$$

Substituting (17) and (18) into the constraint equation in (13) yields the Lagrange multiplier vector as

$$\boldsymbol{\beta}(n)=2\left(\boldsymbol{X}^H(n)\boldsymbol{X}(n)+\boldsymbol{X}^T(n)\boldsymbol{X}^*(n)\right)^{-1}\left(e(n)-\boldsymbol{c}(n)\right) \tag{19}$$

Substituting (19) into (15) and (16), and introducing a small constant $\delta$, it can arrive that

$$\boldsymbol{h}(n+1)=\begin{cases}\boldsymbol{h}(n)+\boldsymbol{X}^*(n)\left(\boldsymbol{X}^H(n)\boldsymbol{X}(n)+\boldsymbol{X}^T(n)\boldsymbol{X}^*(n)+\delta\boldsymbol{I}_P\right)^{-1}\left(e(n)-\boldsymbol{c}(n)\right), & |e_1(n)|>\alpha\\ \boldsymbol{h}(n), & \text{otherwise}\end{cases} \tag{20}$$

$$\boldsymbol{g}(n+1)=\begin{cases}\boldsymbol{g}(n)+\boldsymbol{X}(n)\left(\boldsymbol{X}^H(n)\boldsymbol{X}(n)+\boldsymbol{X}^T(n)\boldsymbol{X}^*(n)+\delta\boldsymbol{I}_P\right)^{-1}\left(e(n)-\boldsymbol{c}(n)\right), & |e_1(n)|>\alpha\\ \boldsymbol{g}(n), & \text{otherwise}\end{cases} \tag{21}$$

where $e_1(n)$ is the first element of the estimation error vector $e(n)$.

According to [9, 10], the parameter $\boldsymbol{c}(n)$ is set to:

$$c_l(n)=\begin{cases}\alpha\,\text{sign}\left(e_l(n)\right), & l=1\\ e_l(n), & l=2,3,\ldots,P\end{cases} \tag{22}$$

the final weight update equations of the ASM-APA can be described as

$$\boldsymbol{h}(n+1)=\boldsymbol{h}(n)+\mu(n)e_1(n)\boldsymbol{X}^*(n)\left(\boldsymbol{X}^H(n)\boldsymbol{X}(n)+\boldsymbol{X}^T(n)\boldsymbol{X}^*(n)+\delta\boldsymbol{I}_P\right)^{-1}\boldsymbol{u}_1 \tag{23}$$

$$\boldsymbol{g}(n+1)=\boldsymbol{g}(n)+\mu(n)e_1(n)\boldsymbol{X}(n)\left(\boldsymbol{X}^H(n)\boldsymbol{X}(n)+\boldsymbol{X}^T(n)\boldsymbol{X}^*(n)+\delta\boldsymbol{I}_P\right)^{-1}\boldsymbol{u}_1 \tag{24}$$

where

$$\mu(n)=\begin{cases}1-\dfrac{\alpha}{|e_1(n)|}, & |e_1(n)|>\alpha\\ 0, & \text{otherwise}\end{cases} \tag{25}$$

and $\boldsymbol{u}_1=[1,0,\ldots,0]^T$.

## 4. Performance analysis

### 4.1. Computational complexity

The number of multiplications required for various algorithms at each iteration when the weight vectors are updated and not updated are given in Table I. Here $N$ denotes the length of the filter, and $P$ is the projection order. The ACNLMS and the SM-ACNLMS have an equal number of multiplications and the lowest complexity. The AAPA and the ASM-APA also have equal complexity. But the update conditions of the SMF are restricted, i.e., whether the weight vector is updated or not depends on (30). Thus, the average number of multiplications of the algorithm per iteration is defined as

$$M_{av}=F_{up}M_{up}+(1-F_{up})M_{nup} \tag{26}$$

where $F_{up}$ is the average update rate (UR), $M_{up}$ is the number of multiplications while updating and $M_{nup}$ is the number of multiplications while not updating with $M_{up} \gg M_{nup}$. Furthermore, the average UR of the AAPA is 1 while the average UR of the ASM-APA is usually less than 0.5. Consequently, the average number of multiplications for the ASM-APA is much less than for the AAPA at each iteration, which illustrates that the ASM-APA is capable of reducing the computational effort. And the conclusions will be confirmed in computer simulations.

TABLE I

Computational complexity

| | Algorithms | Multiplications |
|---|---|---|
| Update | ACNLMS | $5N$ |
| | SM-ACNLMS | $5N$ |
| | AAPA | $(2P^2+4P)N+P^2$ |
| | ASM-APA | $(2P^2+4P)N+P^2$ |
| No update | ACNLMS | / |
| | SM-ACNLMS | $2N$ |
| | AAPA | / |
| | ASM-APA | $2PN$ |

**4.2. Stability Analysis**

In this subsection, the stability of the proposed ASM-APA will be discussed. Observing (23)-(25), ASM-APA can be regarded as a variable step size AAPA. So, the convergence range of $\mu$ for the ASM-APA will be derived first.

The two updating equations (23) and (24) may be associated by

$$\boldsymbol{w}(n+1)=\boldsymbol{w}(n)+\mu\boldsymbol{U}(n)\boldsymbol{Z}(n)e(n) \tag{27}$$

where $\boldsymbol{w}(n)=\begin{bmatrix}\boldsymbol{h}(n)\\ \boldsymbol{g}(n)\end{bmatrix}$ is the augmented matrix of $\boldsymbol{h}$ and $\boldsymbol{g}$, $\boldsymbol{U}(n)=\begin{bmatrix}\boldsymbol{X}^*(n)\\ \boldsymbol{X}(n)\end{bmatrix}$ is the augmented input matrix, and

$$\boldsymbol{Z}(n)=\left(\boldsymbol{X}^H(n)\boldsymbol{X}(n)+\boldsymbol{X}^T(n)\boldsymbol{X}^*(n)+\delta\boldsymbol{I}_P\right)^{-1} \tag{28}$$

Subtract the augmented form of the optimal weight vectors $\boldsymbol{w}_o=\begin{bmatrix}\boldsymbol{h}_o\\ \boldsymbol{g}_o\end{bmatrix}$ from (27) and take expectations, it can be received that

$$E\left[\tilde{\boldsymbol{w}}(n+1)\right]=E\left[\tilde{\boldsymbol{w}}(n)\right]-\mu E\left[\boldsymbol{U}(n)\boldsymbol{Z}(n)e(n)\right] \tag{29}$$

where $\tilde{\boldsymbol{w}}(n)=\boldsymbol{w}_o-\boldsymbol{w}(n)$ denotes the weight error vector.

To facilitate analysis, suppose the noise $v(n)$ is *i.i.d* with zero mean and independent of $\boldsymbol{U}(n)$. Moreover, it is supposed that $\tilde{\boldsymbol{w}}(n)$ is independent of $\boldsymbol{U}(n)\boldsymbol{Z}(n)\boldsymbol{U}^H(n)$. Therefore, (29) is able to be calculated as

$$E\left[\tilde{\boldsymbol{w}}(n+1)\right]=\left(\boldsymbol{I}-\mu E\left[\boldsymbol{U}(n)\boldsymbol{Z}(n)\boldsymbol{U}^H(n)\right]\right)E\left[\tilde{\boldsymbol{w}}(n)\right] \tag{30}$$

The convergence range of $\mu$ is

$$0 < \mu < \frac{2}{\lambda_{\max}\left(E\left[\varpi(n)\right]\right)} \tag{31}$$

where $\varpi(n) = \boldsymbol{U}(n)\boldsymbol{Z}(n)\boldsymbol{U}^H(n)$, and $\lambda_{\max}$ is the maximum eigenvalue of $E\left[\varpi(n)\right]$.

While $\delta$ is small enough, the convergence range is approximate to $(0,2)$. And the step size defined by (30) always changes in $(0,1)$. Consequently, the proposed algorithm is stable.

## 5. Computer Simulations

The computational simulation experiments of the ASM-APA are presented in this section. It is applied to linear system identification and echo cancellation to test the performance of the ASM-APA. The MSE is an indicator for evaluating the capability of the ASM-APA, and it is calculated by

$$\text{MSE}{=}10*\log_{10}\left(\|e(n)\|^2\right) \tag{32}$$

TABLE II

The update rate of the ASM-APA with different $\alpha$

| $\alpha$ | UR |
|---|---|
| $\sqrt{\sigma_v^2}$ | 70.50% |
| $\sqrt{2\sigma_v^2}$ | 56.27% |
| $\sqrt{5\sigma_v^2}$ | 30.73% |
| $\sqrt{10\sigma_v^2}$ | 14.00% |
| $\sqrt{15\sigma_v^2}$ | 8.77% |

### 5.1. System Identification

In this part, we will analyze the connections of the parameters with the performance and compare the proposed ASM-APA with other algorithms in different experimental settings. The input signal $\boldsymbol{x}(n) = [x(n), x(n-1), \ldots, x(n-N+1)]^T$ is the complex-valued Gaussian random signal. The fourth-order autoregressive model then produces:

$$x(n) = 0.95x(n-1) + 0.3x(n-2) + 0.1x(n-3) - 0.5x(n-4) \tag{33}$$

The unknown weight vectors $\boldsymbol{h}_o$ and $\boldsymbol{g}_o$ obey the zero-mean Gaussian distribution with the variance 0.25 and are randomly generated, $\delta = 10^{-5}$. The background noise is complex-valued Gaussian white noise, and the variance with both the parts of real and imaginary is $\sigma_v^2$. Define the update rate of the ASM-APA as $\text{UR} = C_{\text{up}} / L / run$, where $C_{\text{up}}$ is the update times of the weight vectors, $L$ represents the total iterations, and $run$ is the number of the algorithm runs. All simulation results are obtained by averaging 100 experiments and then low-pass filtering except for the echo cancellation experiment.

Fig.1 shows the MSE curves of the ASM-APA with different error bounds $\alpha$ . Table II corresponds to Fig.1 and shows the update rates of the weight vectors, in which the data on update rates are taken to two decimal places. $P$ is set to be 4, and the upper error limits are $\alpha = \sqrt{a\ \sigma_v^2}$ , $a = 1,2,5,10,15$. As can be caught in Fig.1 that the ASM-APA converges slightly faster, but the steady-state error is larger when $a = 1,2$. While $a = 5,10,15$, the algorithm converges a little slower, but with less steady-state error. Meanwhile, it can be found from Table II that as $a$ takes on larger values, the update rate becomes lower, and all are smaller than the AAPA. Even when $a$ is 5 and greater, the update rate is less than half that of the AAPA. In summary, the selection of $a$ has a hold on the convergence performance of the ASM-APA and the average update rate of the weight vectors.

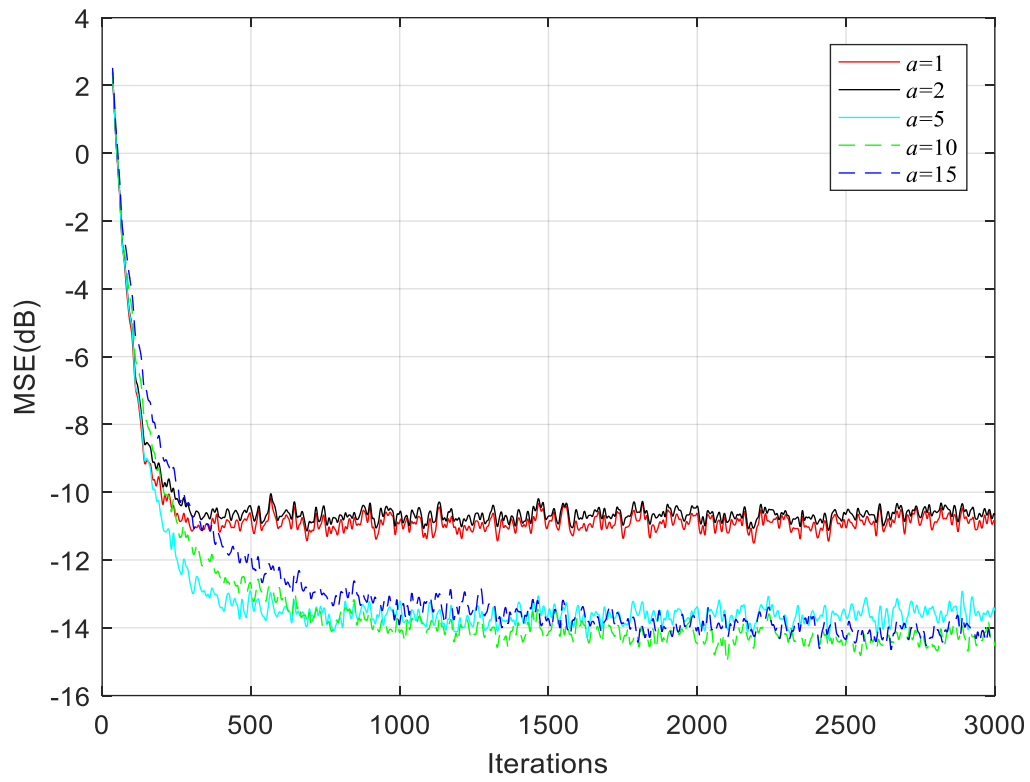

**Fig. 1.** MSE curves of the ASM-APA with different $a$ . [ $N=11, \sigma_v^2 = 0.01, P = 4$ ].

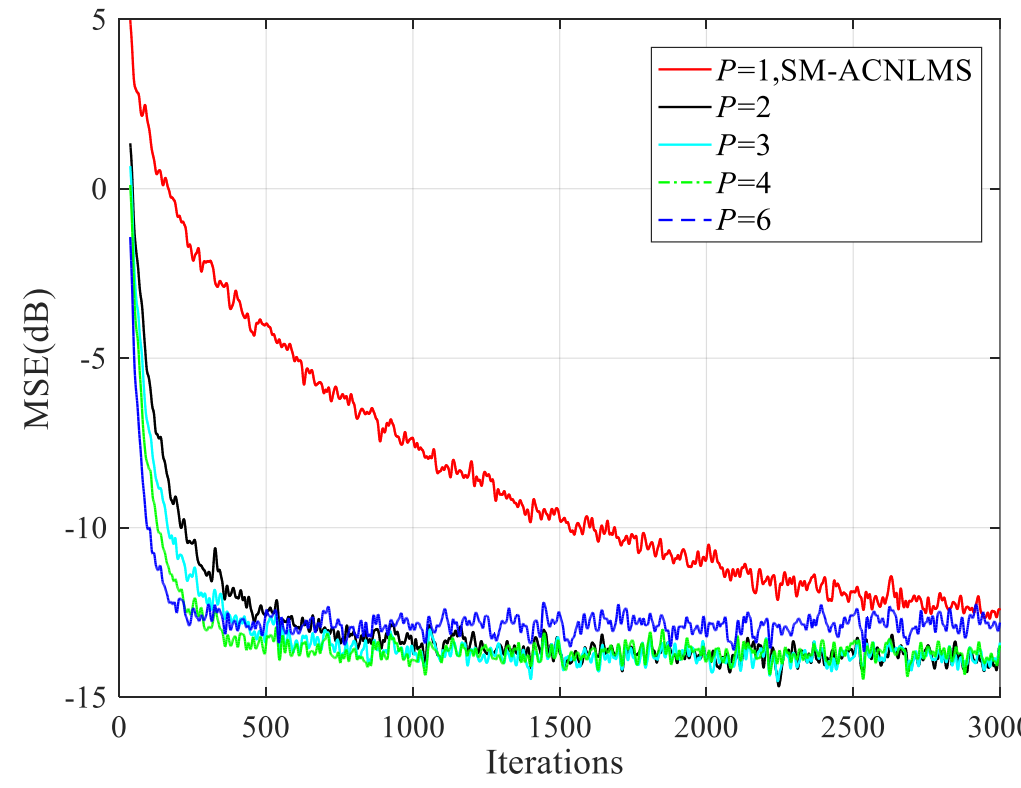

**Fig. 2.** MSE of the ASM-APA with different $P$ . [ $N=11, \sigma_v^2 = 0.01, \alpha = \sqrt{5\sigma_v^2}$ ].

TABLE III

The update rate and complexity of the ASM-APA with different $P$

| $P$ | UR | Average multiplications |
|---|---|---|
| 1 | 49.63% | 44 |
| 2 | 45.00% | 105 |
| 3 | 29.13% | 146 |
| 4 | 27.97% | 216 |
| 6 | 31.13% | 431 |

Fig.2 plots the MSE curves of the ASM-APA with $P=1,2,3,4,6$, and $\alpha=\sqrt{5\sigma_v^2}$. The update rate and the complexity are enumerated in Table III. The complexity is mainly illustrated by the average number of multiplications, and the data in Table III are rounded. When $P=1$, the ASM-APA degenerates into SM-ACNLMS. From Fig.2, it may be known the convergence speed of the ASM-APA gradually accelerates with the increases of $P$. In the same time, the steady-state errors of all the curves of the ASM-APA are approximately equivalent with $P=2,3,4$. And when $P=6$, the steady-state misalignment of the ASM-APA becomes larger. What is more, the update rate progressively decreases and rises for $P=6$. Meanwhile, the average number of multiplications of the algorithm has been always growing.

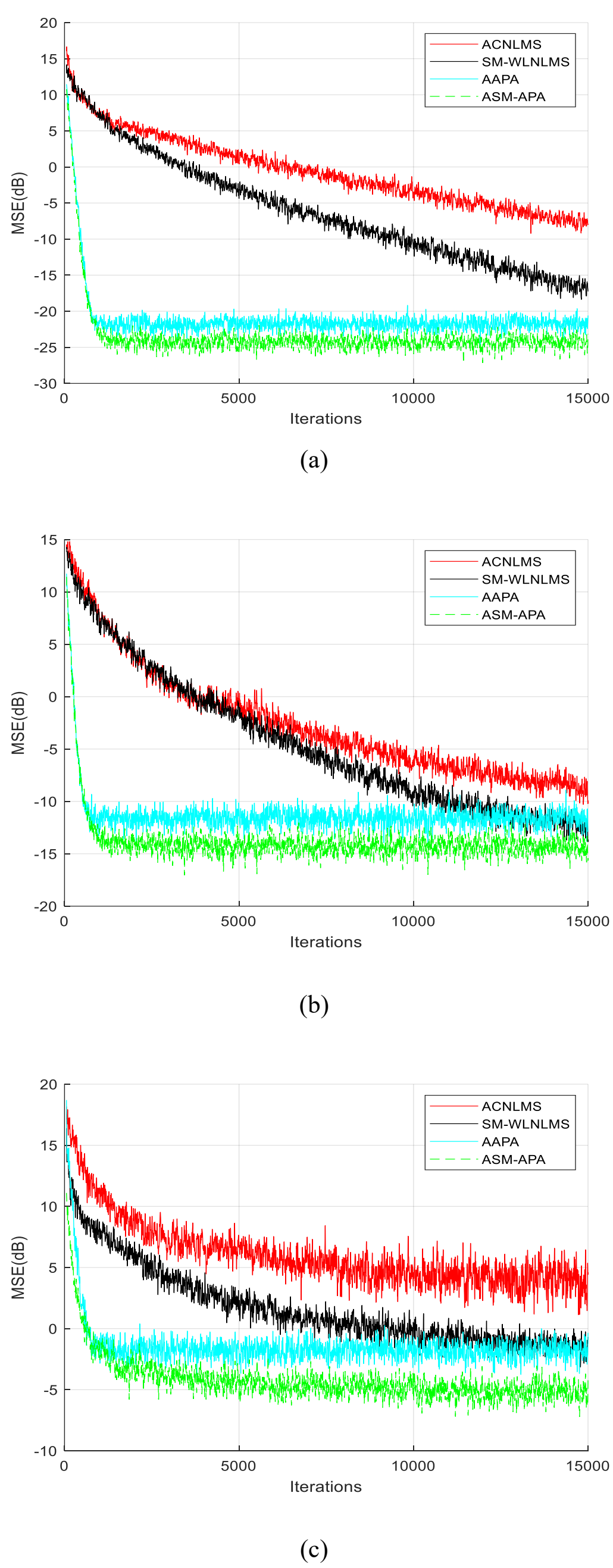

**Fig. 3.** Comparison of MSE with ACNLMS, SM-ACNLMS, AAPA and the proposed ASM-APA under different noise variance $\sigma_v^2$: (a) $\sigma_v^2=0.001$; (b) $\sigma_v^2=0.01$; (c) $\sigma_v^2=0.1$. [ $N=32, P=4, \alpha=\sqrt{5\sigma_v^2}$ ].

TABLE IV

The update rate and complexity of various algorithms under different variance

| Variance | *P* | UR | Average multiplications |
|---|---|---|---|
| 0.001 | ACNLMS | 100% | 160 |
| | SM-ACNLMS | 99.00% | 159 |
| | AAPA | 100% | 1552 |
| | ASM-APA | 37.70% | 745 |
| 0.01 | ACNLMS | 100% | 160 |
| | SM-ACNLMS | 96.40% | 157 |
| | AAPA | 100% | 1552 |
| | ASM-APA | 34.60% | 704 |
| 0.1 | ACNLMS | 100% | 160 |
| | SM-ACNLMS | 82.40% | 143 |
| | AAPA | 100% | 1552 |
| | ASM-APA | 32.40% | 676 |

In Fig.3, the MSE curves of ACNLMS, SM-ACNLMS, AAPA and the ASM-APA with different noise variances $\sigma_v^2$ are present. The $P$ of AP-type algorithms is 4, and the error bound $\alpha$ of SM-type algorithms is $\sqrt{5\sigma_v^2}$ . The step sizes of ACNLMS and AAPA are 0.4 and 0.7 respectively. From Fig.3 it can be concluded that the AP-type algorithms converge at much higher speed than the NLMS for the input signals with high correlation. And as the noise variance increases, the ASM-APA converges from marginally faster than the AAPA to similar to the AAPA, but the steady-state error of the ASM-APA always smaller than that of the AAPA. In addition, Table IV illustrates that the ACNLMS and the SM-ACNLMS have the lowest computational complexity with an average of less than 200 multiplications. The ASM-APA has a consistent update rate of 30%-40%, which is only one third of that of the AAPA. Overall, the ASM-APA reduces the updates of the weight vectors while maintaining a low steady-state misalignment, effectively decreasing the complexity.

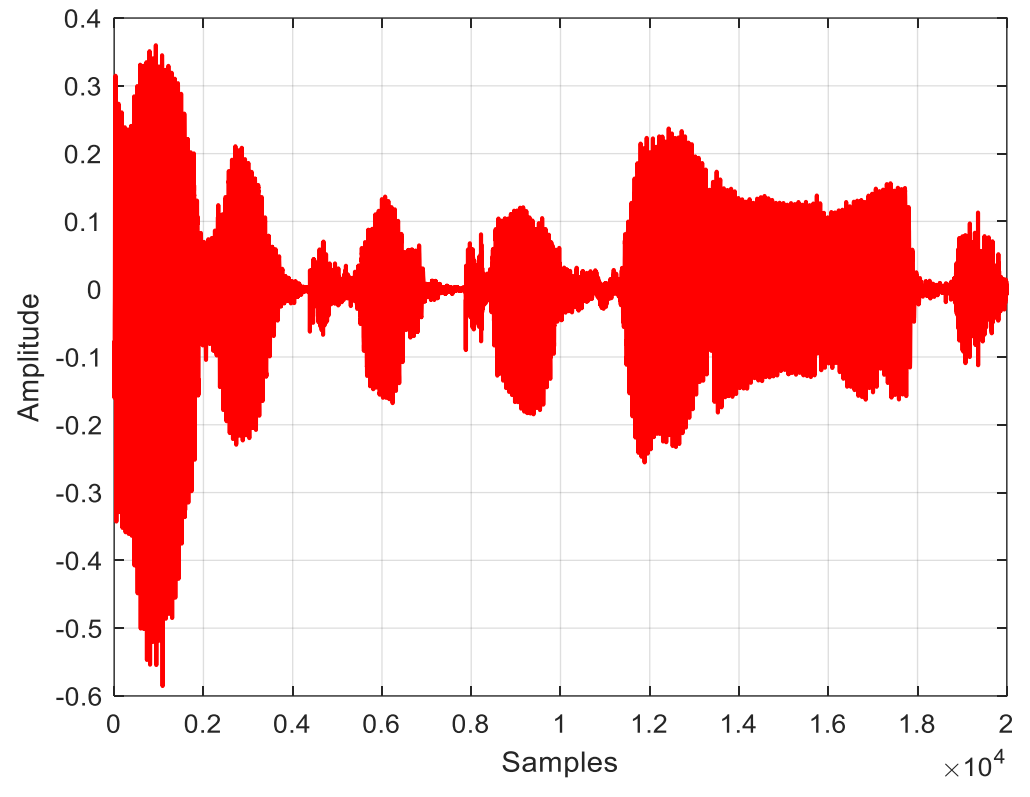

**Fig. 4.** Speech signal.

## 5.2 Stereophonic acoustic echo cancellation

The proposed ASM-APA will be tested in the real system using stereophonic acoustic echo cancellation (SAEC). The Normalized Mean-Squared-Deviation (NMSD) is adopted to assess the performance of the ASM-APA, which is designed as

$$\mathrm{NMSD} = 10\log_{10}\left(\frac{\|\boldsymbol{w}_{\mathrm{o}} - \boldsymbol{w}(n)\|^2}{\|\boldsymbol{w}_{\mathrm{o}}\|^2}\right) \tag{34}$$

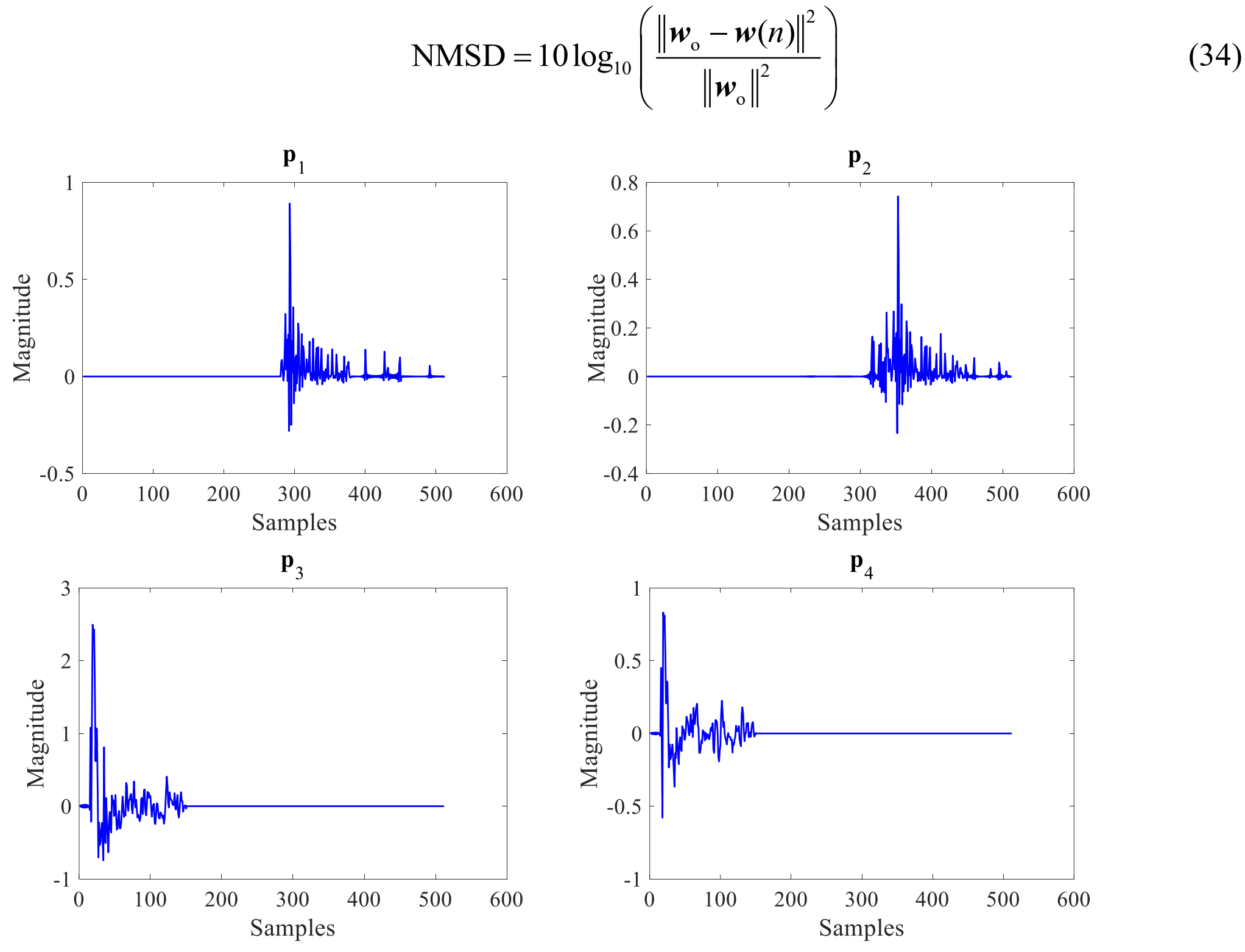

**Fig. 5.** Acoustic echo channel $\{\boldsymbol{P}_1, \boldsymbol{P}_2, \boldsymbol{P}_3, \boldsymbol{P}_4\}$.

The input is a real speech signal as shown in Fig.4. After passing through two different finite impulse response filters, the input signal $\boldsymbol{x}(n)$ are designed as the outputs of the two filters, which can be described as $\boldsymbol{x}(n) = \boldsymbol{x}_1(n) + j\boldsymbol{x}_2(n)$. The half-wave rectifiers are then used to correct the non-linearities of real and imaginary parts separately with a parameter of 0.5. Moreover, there are four echo channels $\{\boldsymbol{P}_1, \boldsymbol{P}_2, \boldsymbol{P}_3, \boldsymbol{P}_4\}$ in the SAEC with a length of 512, and the impulse response is drawn in Fig.5. The echo signal is modelled as

$$y(n) = \boldsymbol{x}^T(n)\boldsymbol{h}_t + \boldsymbol{x}^H(n)\boldsymbol{g}_t + v(n) \tag{35}$$

and $\boldsymbol{h}_t$ and $\boldsymbol{g}_t$ are described by

$$\boldsymbol{h}_t = \frac{(\boldsymbol{p}_1 + \boldsymbol{p}_4)}{2} - j\frac{(\boldsymbol{p}_3 - \boldsymbol{p}_2)}{2} \tag{36}$$

$$\boldsymbol{g}_t = \frac{(\boldsymbol{p}_1 - \boldsymbol{p}_4)}{2} + j\frac{(\boldsymbol{p}_3 + \boldsymbol{p}_2)}{2} \tag{37}$$

and the noise $v(n)$ in the proximal signal is the Gaussian white noise with the variance of 0.01.

Fig.6 compares the NMSD of the AAPA with $\mu = 0.0003$ and the ASM-APA with its error bound $\alpha = 1.5$. As can be observed, the ASM-APA has achieved faster convergence and lower steady-state inaccuracy with an update rate of only 0.52% than the AAPA. Thus, the ASM-APA has both better performance and smaller computational complexity.

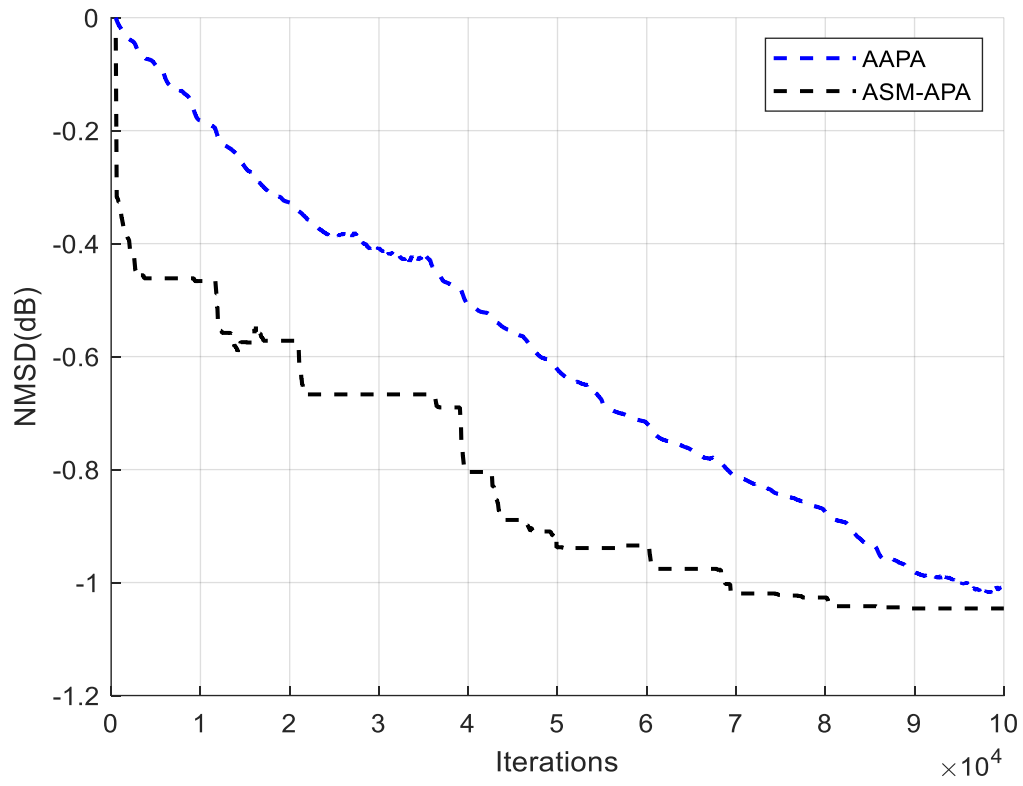

**Fig. 6.** NMSD curves of AAPA and the ASM-APA. [ $P=4$ , AAPA: $\mu=0.0003$ , ASM-APA: $\alpha=1.5$ ].

## 6. Conclusion

The set-membership augmented affine projection algorithm has been proposed in accord with data selectivity in this letter. The algorithm creates an optimal feasible space for the updates. The weight vectors update only when the errors exceed the threshold, which effectively reduces the computational complexity the ASM-APA. The complexity and the stability have also been presented. In addition, the computational simulations of the identification of the unknown system and stereophonic acoustic echo cancellation have been carried out. It has turned out that the proposed ASM-APA has comparable performance for adaptive filtering with the colored and non-circular input signals.

## Acknowledgments

This work was somewhat supported by National Natural Science Foundation of China (grant: 62171388, 62001183, 61871461, 61571374), Jiangsu Province Key Research and Development Plan Modern Agriculture Project (BE2023345), Suqian Sci&Tech Program (M202305).

## Data availability

The data that support the findings of this study are available from the corresponding author on request.

## Publisher's Note